\documentclass{svjour3}
\usepackage{graphicx}
\journalname{General Relativity and Gravitation}
\begin{document}

\title{Gravitational wave detection with single-laser atom interferometers}
\author{Nan Yu \and Massimo Tinto}
\institute{Nan Yu \at
              Jet Propulsion Laboratory, MS 298\\
              4800 Oak Grove Drive,\\
              Pasadena, CA 91109\\
              Tel.: +1-818-393-4093\\
              Fax: +1-818-393-6773\\
              \email{Nan.Yu@jpl.nasa.gov}
              \and
              Massimo Tinto \at
              Jet Propulsion Laboratory, MS 238-737\\
              4800 Oak Grove Drive,\\
              Pasadena, CA 91109\\
              Tel.: +1-818-354-0798\\
              Fax: +1-818-354-2825\\
              \email{Massimo.Tinto@jpl.nasa.gov}
}

\date{Received: date / Accepted: date}
\maketitle

\begin{abstract}
  
  We present a new general design approach of a broad-band detector of
  gravitational radiation that relies on two atom interferometers
  separated by a distance $L$.  In this scheme, only one arm and one
  laser will be used for operating the two atom interferometers. We
  consider atoms in the atom interferometers not only as perfect
  inertial reference sensors, but also as highly stable clocks. Atomic
  coherence is intrinsically stable and can be many orders of
  magnitude more stable than a laser. The unique one-laser
  configuration allows us to then apply time-delay interferometry to
  the responses of the two atom interferometers, thereby canceling the
  laser phase fluctuations while preserving the gravitational wave
  signal in the resulting data set. Our approach appears very
  promising.  We plan to investigate further its practicality and
  detailed sensitivity analysis.


\end{abstract}

\section{Introduction}
\label{intro}

The detection of gravitational radiation is one of the most
challenging efforts in physics in this century. A successful
observation will not only represent a great triumph in experimental
physics, but will also provide a new observational tool for obtaining
better and deeper understandings about its sources, as well as a
unique test of the proposed relativistic theories of gravity
\cite{Thorne1987}.

Non-resonant detectors of gravitational radiation (with frequency
content $0 < f < f_0$) have one or more arms with coherent trains of
electromagnetic waves (of nominal frequency $\nu_0 \gg f_0$), or
beams, and at points where these intersect, relative fluctuations of
frequency or phase are measured (homodyne detection).  Frequency
fluctuations in a narrow Fourier band can alternatively be described
as fluctuating sideband amplitudes. Interference of two or more beams,
produced and monitored by a (nonlinear) device such as a photo
detector, exhibits these sidebands as a low frequency signal again
with frequency content $0 < f < f_0$.  The observed low frequency
signal is due to frequency variations of the sources of the beams
about $\nu_0$ along their propagation paths, including motions of the
sources and any mirrors (or amplifying microwave or optical
transponders) that do any beam folding, to temporal variations of the
index of refraction along the beams, and, according to general
relativity, to any time-variable gravitational fields present, such as
the transverse traceless metric curvature of a passing plane
gravitational wave train.  To observe these gravitational fields in
this way, it is thus necessary to control, or monitor, all other
sources of relative frequency fluctuations, and, in the data analysis,
to optimally use algorithms based on the different characteristic
interferometer responses to gravitational waves (the signal) and to
other sources (the noise).

In present single-spacecraft microwave Doppler tracking observations,
for instance, many of the noise sources can be either reduced or
calibrated by implementing appropriate frequency links and by using
specialized electronics, so the fundamental limitation is imposed by
the frequency fluctuations inherent to the reference clock that
controls the microwave system.  Hydrogen maser clocks, currently used
in Doppler tracking experiments, achieve their best performance at
about $1000$ seconds integration time, with a fractional frequency
stability of a few parts in $10^{-16}$.  This is the reason why these
one-arm interferometers in space are most sensitive to millihertz
gravitational waves.  This integration time is also comparable to the
microwave propagation (or "storage") time $2 L/c$ to spacecraft en
route to the outer solar system ($L \simeq 3 AU$), so these one-arm,
one-bounce, interferometers have near-optimum response to
gravitational radiation, and a simple antenna pattern.

By comparing phases of split beams propagated along equal but
non-parallel arms, common frequency fluctuations from the source of
the beams can be removed directly and gravitational wave signals at
levels many orders of magnitude lower can be detected. Especially for
interferometers that use light generated by presently available
lasers, which have frequency stability of roughly a few parts in
$10^{-13}/\sqrt{Hz}$ to $10^{-18}/\sqrt{Hz}$ (in the millihertz and
kilohertz bands respectively) it is essential to be able to remove
these fluctuations when searching for gravitational waves of
dimensionless amplitude less than $10^{-20}$ in the millihertz band
\cite{PPA98,TA_1998}, or down to $10^{-21} - 10^{-23}$ desired in the
kilohertz frequency band \cite{LIGO}.  Combined with the fact that
plane gravitational waves have a spin-two polarization symmetry, this
implies that the customary right-angled Michelson configuration is
optimal.  The response to gravitational waves is then maximized in
Earth-based systems by having many bounces in each arm
\cite{Weiss,Drever}.

The frequency band in which a ground-based interferometer can be made
most sensitive to gravitational waves \cite{Thorne1987} ranges from
about a few tens of Hertz to about a few kilohertz, with arm lengths
ranging from a few tens of meters to a few kilometers.  Space-based
interferometers, such as the coherent microwave tracking of
interplanetary spacecraft \cite{Armstrong2006} and the proposed Laser
Interferometer Space Antenna (LISA) mission \cite{PPA98} are most
sensitive to millihertz gravitational waves and have arm lengths
ranging from $10^6$ to $10^8$ kilometers.

Recently, atom-wave interferometers have been proposed as potentially
new gravitational waves detectors, as their technology has reached a
high-level of maturity in providing extremely sensitive inertial
sensors \cite{Kasevich}. The discussions on how to use atom
interferometers as gravitational wave detectors have mainly focused on
two fundamental types of proposed approaches. The first relies on
exploiting the atom wave and the interferometers directly
\cite{Chiao,Speliotopoulos}. In this scheme, atoms in an atom-wave interferometer
correspond to photons in a Michelson laser interferometer, and the
effects of a gravitational wave signal are measured by monitoring the
phase (momentum) changes of the atoms due to the GW. Although this
design has been shown of not providing any sensitivity improvements
over optical interferometers \cite{Misner}, it stimulated more
thoughts on the subject. The result was an alternative design in which
atom interferometers (AI) are used as local inertial sensors and laser
beams (which are used for operating the AIs) imprint on the atoms the
phase fluctuations generated by a gravitational wave signal
propagating across the detector \cite{Dimopoulos_etal2008}.  This
approach is fundamentally the same as using AI for gravity gradient
measurements, which has been pursued both on the ground and in space.

Along the line of using atoms as test masses in a gravitational wave
detector, we present in this paper an alternative configuration in
which atoms are treated as local free-falling {\underline {clocks}}.
Consider a detector configuration with two ensembles of atoms
separated by a distance $L$, in which only a {\underline {single}}
laser beam is used to operate them.  The laser interrogates the atoms
similar to a local oscillator laser interacting with atoms in an
optical clock. The results give the phase differences between the
laser and the highly coherent atomic internal oscillations. As the laser
phase fluctuations enter into the responses of the two phase
difference measurements at times separated by the one-way-light-time,
$L$ (units in which the speed of light $c = 1$), we show that the
laser phase fluctuations can be exactly canceled (while retaining the
gravitational wave signal) by applying time-delay interferometry (TDI)
\cite{TD2005} to the phase measurement data.

The rest of the paper is organized as follows. In Section \ref{SecII}
we consider a detector configuration with two AIs separated by a
distance $L$, in which only a laser beam is used to operate them.  In
Section \ref{SecIII} we turn to the problem of canceling the laser
phase fluctuations by using the data from the two AIs. Since the laser
phase noise enters into the responses of the two AIs at times
separated by the one-way-light-time $L$ (units in which the speed of
light $c = 1$), we show that it can exactly be canceled (while
retaining the gravitational wave signal) by applying time-delay
interferometry (TDI) \cite{TD2005} to the AIs data. In Sections
\ref{SecIV} we make some further considerations about how to enhance
the sensitivity of these single-laser AIs by large-momentum-transfer
techniques, while in Section \ref{SecV} we present our conclusions and
remarks on future prospects for this gravitational wave detector
design.

\section{Laser phase measurements with atom interferometry}
\label{SecII}

The fundamental limitation of a one-arm Doppler measurement
configuration (such as that of interplanetary spacecraft tracking
experiments) is determined by the frequency stability of the ``clock''
that defines the frequency of the electromagnetic link. The most
stable clocks are presently optical atomic clocks. Optical clocks have
already shown stabilities of about $10^{-17}$ over $1000$ sec
integration time \cite{Clocks_HighlyStable}. This is accomplished by
frequency-locking a highly stabilized laser to an atomic transition as
an ideal passive frequency standard. The intrinsic atomic coherence is
only limited by its natural lifetime.  External perturbations cause
additional frequency fluctuations, which may be controlled to a level of
$10^{-18}$ and lower \cite{MadjeClock,YuClock}.

\begin{figure}
 \begin{center}
  \includegraphics[width=5.0in, angle = 0.0]{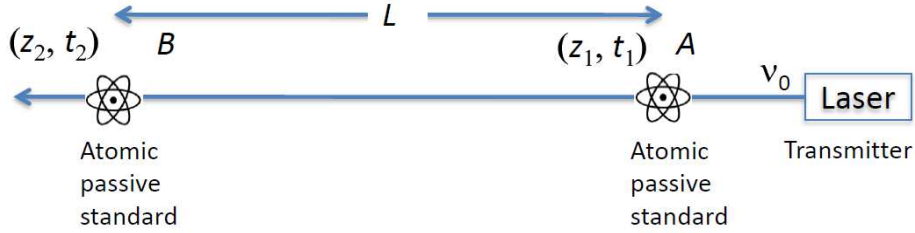}
   \end{center}
    \caption{A laser of nominal frequency $\nu_0$ shines on two AIs,
      separated by a distance $L$, at space-times events $(z_1, t_1)$
      and $(z_2, t_2)$.}
\label{Fig1}
\end{figure}

These considerations imply that we might use atoms directly as ideal
local reference oscillators for gravitational wave detection. To
illustrate this point, let's assume to have two identical atomic
systems that are free-falling. The two atomic systems, A and B, are
placed $L$ light-seconds apart.  A coherent laser is near the atomic
system A, as shown in Fig. \ref{Fig1}.
Like in a conventional atomic clock, the laser excites the atomic
systems through a Rabi (or Ramsey) interrogation method. In essence,
the laser frequency and phase are compared to that of the atomic
internal clock, i.e.  the oscillation between two atomic energy levels
of the atom.  In Figure \ref{Fig1} the laser shines from the
right-hand-side of our proposed experimental setup, its field
interacts with the atomic system A at time $t_1$ (which has an
interrogation time $T$), and then it reaches the atomic system B (that
has the same interrogation time as A) at time $t_2$ ($L$ seconds
later).  This results in two phase differences containing the laser
phase noise at space-times ($z_1, t_1$) and ($z_2, t_2$) and noises of
the local atomic coherence over the interrogation time $T$. We will
show in Section \ref{SecIII} that (i) the laser phase noise can be
canceled exactly by applying TDI \cite{TD2005} to the data measured by
the two AIs, and (ii) the resulting new data set retains sensitivity
to the gravitational wave signal and is now limited by the stability
of the passive atomic clocks, which can be several orders of magnitude
smaller than that of the laser.

It is important to point out here that making a ``stand-alone'' atomic
clock requires a local oscillator (typically a laser) whose stability
then becomes the main limitation of the atomic standard stability over
the interrogation time.  Currently, the best approach is to use as
local oscillator a laser that is frequency locked to a highly
stabilized Fabry-Perot cavity. This approach however has already
reached its fundamental limit due to thermal noise \cite{YeCavity}.
Our approach of coupling two passive atomic standards with the same
laser completely circumvents the laser stability requirement. It is in
fact easy to see that the laser phase noise cancellation is valid even
in the large phase noise limit where the phase noise of the laser
fluctuates more than one radian during the interrogation time $T$.

\section{Single-laser atom interferometers as free-falling test masses}
\label{SecIII}

One of the key requirements in interferometric gravitational wave
experiments is for the local reference frames to be as much inertially

free as possible.  This is to reduce any none gravitational forces and
local gravitational disturbances that can cause changes in the laser
phase. Ground-based interferometers achieve a high-level of seismic
isolation of their mirrors by using either passive or active isolation
systems \cite{LIGO,VIRGO}.  Space-based detectors instead, such as
LISA, achieve inertial isolation by using highly sophisticated
drag-free test masses.  Although in principle one could trap atoms in
such test masses, it is more practical to rely on laser-cooled atoms
in ultra-high vacuum as alternative drag-free test masses and directly
use them as reference sensors.

Although an ensemble of laser-cooled free atoms would still have a
finite velocity distribution that would degrade the fringe contrast
because of inhomogeneous Doppler dephasing, there exist techniques
(such as spin echo, for instance) that allow one to compensate for
inhomogeneous dephasing effect. This is in fact discussed in relation
with Doppler-free Ramsey-Borde interferometers \cite{Ramsey_Borde},
which have been successfully used as optical clocks with free-falling
atoms \cite{CaClock}. Since this technique requires a pair of
counter-propagating laser beams, it cannot be applied to our proposed

design since the phase fluctuations from the two lasers cannot be
canceled both at the same time.

\begin{figure}
  \begin{center}
   \includegraphics[width=5.0in, angle = 0.0]{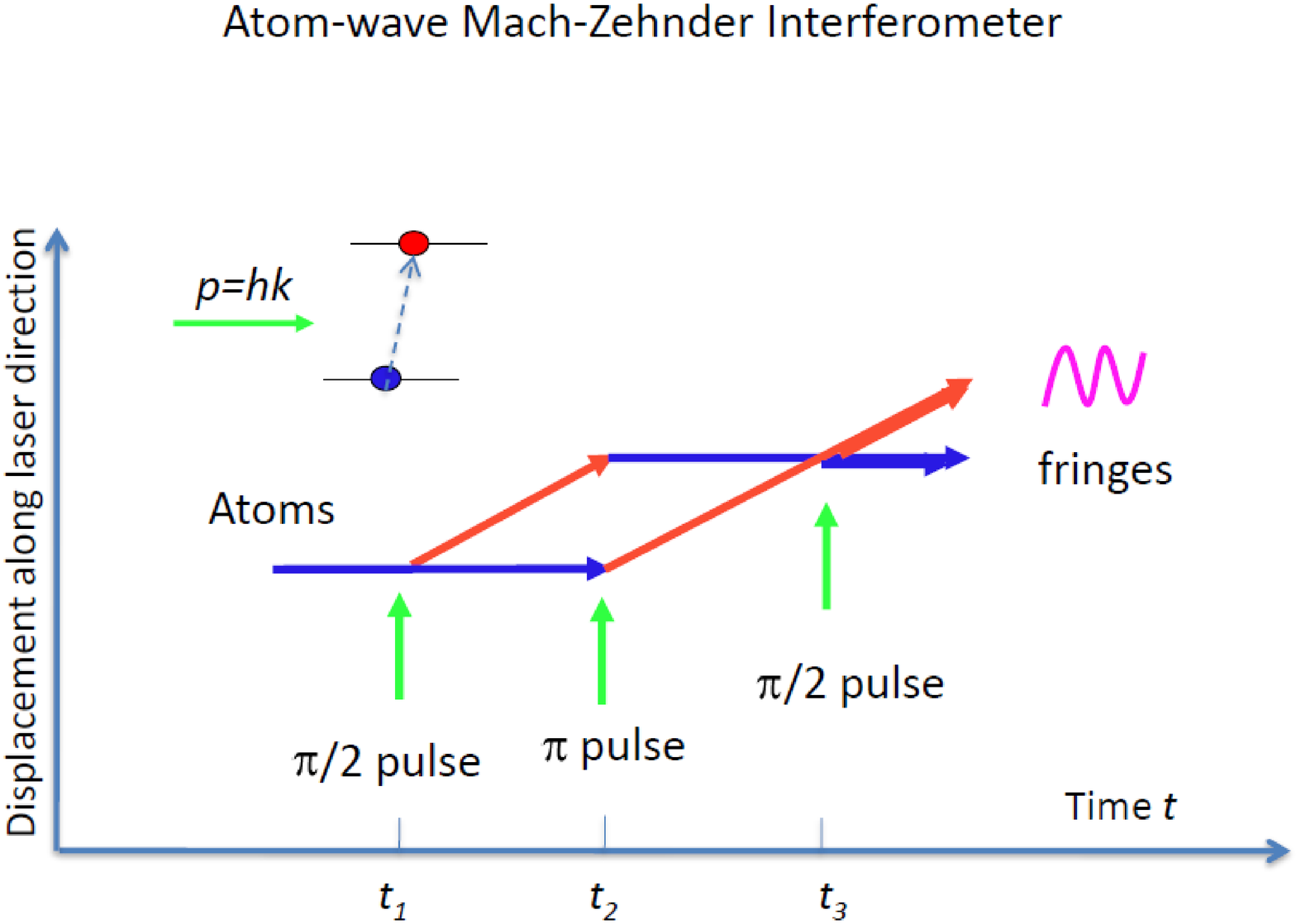}
    \end{center}
     \caption{Single laser light pulse atom interferometer through an optical transition
     }
\label{Fig2}
\end{figure}

To overcome this problem, recall that, in an atom interferometer with
$\frac{\pi}{2} - \pi - \frac{\pi}{2}$ stimulated Raman pulse sequence
\cite{Kasevich}, the resulting phase shift measured by the atom
interferometer is equal to $\Delta \Phi = k_{eff} \ a \ T^2 \ + \ \phi
(t_1) \ - \ 2 \phi (t_2) \ + \ \phi (t_3)$, where $k_{eff}$ is the
effective wave-number of the Raman lasers, $a$ is the acceleration of
the atoms, and $\phi$ is the phase difference of the two Raman lasers.
This result is independent of the individual atom velocities. All
$\frac{\pi}{2} - \pi - \frac{\pi}{2}$ type atom interferometers
considered so far use stimulated Raman transitions with a pair of
counter-propagating laser beams.  However, this lack of velocity
independence is not due to the presence of counter-propagating beams,
but results from the $\pi$ phase reversal taking place at the middle
of the interferometer, quite similarly to spin echo in nuclear
magnetic resonance \cite{NMR}. In fact, it can be shown that one can
achieve a $\frac{\pi}{2} - \pi - \frac{\pi}{2}$ sequence in an atom
interferometer with only one laser from one direction only, as
sketched in Fig. \ref{Fig2}. In this case, the two atomic levels
involved will be an optical clock transition.  The resulting phase of
the optical transition based atom interferometer will be exactly the
same as in stimulated Raman transitions: $\Delta \Phi = k a T^2 \ + \ 
\phi (t_1) \ - \ 2 \phi (t_2) \ + \ \phi (t_3)$, where now $k$ is the
wave number and $\phi$ is the phase of the laser.

For an ensemble of free-falling atoms the measured interferometer
phase is equal to $\Delta \Phi = \phi (t_1) - 2 \phi (t_1 + T) + \phi
(t_1 + 2T)$. This expression indicates that such an atom
interferometer can compare the phase of the laser against the atomic
internal clock coherence.  This phase difference measurement can then
be used as part of our proposed gravitational wave detector design, as
we mathematically show below.

\subsection{The gravitational wave signal}

\begin{figure}
 \begin{center}
  \includegraphics[width=4.0in, angle = 0.0]{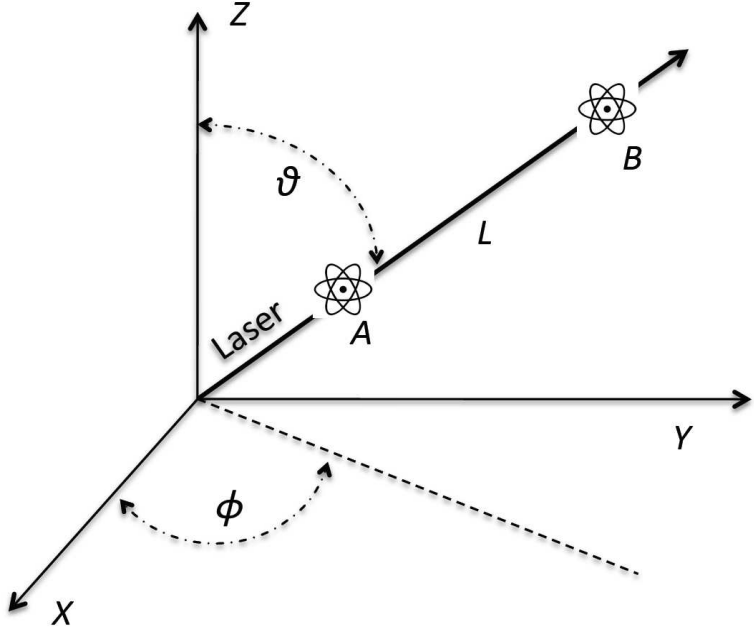}
   \end{center}
    \caption{Coherent laser light is transmitted along arm $L$ from AI
      A to AI B. The plane gravitational wave signal is
      propagating along the $Z$ direction, and the ($X, Y$)
      coordinates are defined in the plane of the wave. In these
      coordinates the direction of propagation of the laser light is
      described by the two polar angles ($\theta, \phi$).}
\label{Fig3}
\end{figure}

Let us assume a plane polarized gravitational wave (GW) propagating
across the experimental setup shown in Fig. \ref{Fig3}. As a result it
will leave an imprint on the phases of the light and the atoms, and
appear as additional phase fluctuations at the output of the two AIs.
The phase fluctuations induced by the GW on the light propagating
along ``the arm'' $L$ of our experimental setup represent the dominant

contribution from the GW to the overall GW phase fluctuations measured
by the atom interferometers \cite{Dimopoulos_etal2008}. This is
because the characteristic size $T$ of the AIs is much smaller than
the arm-length $L$. In what follows, therefore, we will disregard the
effects of the wave on the phase of the atoms.  This means that the
main GW signal will appear at the output of AI B, and the GW signal at
the output of AI A can be assumed to be null.

If we introduce a set of Cartesian orthogonal coordinates ($X, Y, Z$)
in which the wave is propagating along the $Z$-axis and ($X, Y$) are
two orthogonal axes in the plane of the wave (see Figure \ref{Fig3}),
then the frequency of the laser at time $t$, $\nu (t)$, at the beam
splitter of AI B is related to the nominal frequency $\nu_0$ of the
laser and the gravitational wave's amplitudes, $h_+ (t)$ and $h_\times
(t)$, by the following relationship \cite{EW1975,TD2005}

\begin{equation}
\frac{1}{2 \pi \nu_0} \frac{d}{dt} \delta \phi_h (t)
\equiv \frac{\nu (t) - \nu_0}{\nu_0} =   
\frac{1 - \mu}{2} \ \left[ h(t - (1 + \mu)L) \ - \ 
h(t) \right] \ ,
\label{eq:1} 
\end{equation}
where $h(t)$ is equal to 
\begin{equation} 
h (t) = h_+(t) \cos (2 \phi) + h_{\times}(t) \sin (2 \phi) \ ,
\label{eq:1bis}
\end{equation}
and we have denoted with $\delta \phi_h (t)$ the phase fluctuations of
the light due to the GW.  The wave's two amplitudes are defined with
respect to the ($X, Y$) axis, ($\theta, \phi$) are the polar angles
describing the location of AI B with respect to the ($X, Y, Z$)
coordinates, and $\mu$ is equal to $\cos \theta$.

From Eq. (\ref{eq:1}) we deduce that gravitational wave pulses of
duration longer than the one-way-light-time $L$ give a Doppler response
that, to first order, tends to zero with the distance $L$.  This
system essentially acts as a pass-band device, in which the
low-frequency limit $f_l$ is roughly equal to $(L)^{-1}$ Hz, and the
high-frequency limit $f_H$ is set by the atom shot noise in the AIs.

If we assume the arm of the AIs to be equal to $T$, we find that the
phase shifts due to the GW signal at the outputs of the AIs, $R_{1h}
(t)$ and $R_{2h} (t)$, 
are equal to \cite{Dimopoulos2008_2}
\begin{eqnarray}
R_{1h} (t) &=& 0 \ ,
\label{eq:2} 
\\
R_{2h} (t) &=& - [\delta \phi_h (t) - 2 \ \delta \phi_h (t - T) + \delta
\phi_h (t - 2T)] \ .
\label{eq:3} 
\end{eqnarray}
The additional time-signature of the gravitational wave signal in the
response $R_{2h} (t)$ follows from the fact that the light of the
laser affects the atoms of AI B at times $t$, $t - T$, and $t -
2T$; these are the instances when the phase fluctuations of the wave
get imprinted on the phase of the atom.

\subsection{The noise responses}

Since the main goal of this paper is to show that it is possible to
combine the data from the two AIs in such a way to exactly cancel the
phase fluctuations of the laser and retain those from the GW, in this
section we will focus on how the laser phase fluctuations, $P (t)$,
enter into the responses of the two AIs.

As an initial AI laser pulse excites an atom, it establishes the
atomic internal oscillation and imprints its ``instantaneous'' phase
onto the atomic oscillation. While the subsequent phase fluctuation
$P(t)$ of the laser is governed by the laser stability, the atomic
coherence may be perturbed by external environment and can be
characterized by equivalent phase fluctuations $P_i(t), i=1,2$
indicating atoms in different AIs. Subsequent AI laser pulses result
in interference between the laser and the atomic coherence. The
responses of the AIs to the laser phase noise, $P(t)$, and the phase
fluctuations of the radiation emitted by an atom, $P_i (t)$, can be
written in the following form

\begin{eqnarray}
R_{1rad} (t) &=& [P_1(t) - 2 \ P_1(t - T) + P_1(t - 2T)] - 
[P(t) - 2 \ P(t - T) + P(t - 2T)] \
\label{r1rad}
\\
R_{2rad} (t) &=& [P_2 (t) - 2 \ P_2(t - T) + P_2(t - 2T)]
-  [P(t - L) - 2 \ P(t - T - L) 
\nonumber
\\
&& \ + \ P(t - 2T - L)] \ .
\label{r2rad}
\end{eqnarray}
Note that the phase fluctuations of the laser at time $t$ at AI B
were generated $L$ seconds earlier, and for this reason they appear
in $R_{2rad} (t)$ time-delayed by $L$.

By combining Eqs. (\ref{eq:2}, \ref{eq:3}, \ref{r1rad}, \ref{r2rad}),
and denoting with $N_1$, $N_2$ the phase fluctuations due to all other
noises affecting the responses of the two AIs, $R_1 (t)$ and $R_2(t)$,
we get
\begin{eqnarray}
R_{1} (t) &=& [P_1(t) - 2 \ P_1(t - T) + P_1(t - 2T)] - 
[P(t) - 2 \ P(t - T) + P(t - 2T)] 
\nonumber
\\
&& + N_1 (t) \ ,
\label{r1}
\\
R_{2} (t) &=& - [\delta \phi_h (t) - 2 \ \delta \phi_h (t - T) + \delta
\phi_h (t - 2T)] + [P_2 (t) - 2 \ P_2(t - T) + P_2(t - 2T)]
\nonumber
\\
&& -  [P(t - L) - 2 \ P(t - T - L) + P(t - 2T - L)] + N_2 (t) \ .
\label{r2}
\end{eqnarray}

By noticing that the laser phase fluctuations enter into the response
of AI B at a time-shifted by $L$ seconds, it is easy to see that 
the following linear combination of the two responses $R_1$
and $R_2$ cancels them
\begin{eqnarray}
\Delta R(t) &\equiv& R_1(t - L) - R_2(t) = [\delta \phi_h (t) - 2 \
\delta \phi_h (t - T) + \delta \phi_h (t - 2T)] + 
[P_1(t - L) 
\nonumber
\\
&& \ - \ 2 \ P_1(t - T - L) \ + \ P_1(t - 2T - L)] - [P_2 (t) - 2 \
P_2(t - T) + P_2(t - 2T)] 
\nonumber
\\
&& \ + \ N_1 (t - L) - N_2 (t) \ .
\label{AIdiff}
\end{eqnarray}
If we now denote with $\widetilde{\Delta R} (f) $ the Fourier
transform of $\Delta R(t)$, and substitute into it the Fourier
transform of $\delta \phi_h (t)$ from Eq. (\ref{eq:1}), we get the
following expression of the laser noise-free combination
$\widetilde{\Delta R} (f)$ in terms of the remaining noises and the
gravitational wave amplitude $h$ (defined in Eq. (\ref{eq:1bis}) in
terms of the wave's two amplitudes, $h_+ (t)$ and $h_\times (t)$)
\begin{eqnarray}
\widetilde{\Delta R} (f) & = &
\left[ \frac{\nu_0}{i f} \left(\frac{1 - \mu}{2} \right) \ {\widetilde
    h} (f) \ [ e^{2 \pi i f (1 + \mu)L} \ - \ 1 ] \ + \ {\widetilde{P_1}} (f) \
  e^{2 \pi i f L} \ - \ {\widetilde{P_2}} (f) \right] 
\nonumber
\\
&& \times \ [1 - e^{2 \pi i f T}]^2 \ + \ 
{\widetilde{N_1}} (f) \ e^{2 \pi i f L} \ - \ {\widetilde{N_2}} (f)
\label{R}
\end{eqnarray}

Note that the resulting one-arm interferometric response to a GW
signal, given in Eq.  (\ref{R}), is much simpler than that of an
unequal-arm laser interferometer \cite{TD2005}.

\section{Discussion}
\label{SecIV}

We have shown that atoms in their free-falling state can be used in
one-way precision laser interferometry experiments. This is possible
because (i) atomic internal state oscillations can be used as highly
stable clocks for laser phase comparison measurements and, (ii) by
relying on our ``one-way'' configuration, the laser phase fluctuations
can be exactly canceled by applying TDI to the data measured by the
atom interferometers.

It is instructive to look at the sensitivity of our measurement scheme
for gravitational wave detection, since now the fundamental limitation
becomes the phase measurement resolution of the atom interferometers.
In its simplest operational configuration, the phase measurement
resolution is determined by the quantum projection noise of atoms,
commonly known as the atom number shot noise. If $N$ is the total
number of atoms in a AI, the resulting phase noise measurement is
equal $1/\sqrt{N}$.  If we follow the rather optimistic assumption of
operating our AIs with $10^{8} {\rm atoms}/s$
\cite{Dimopoulos_etal2008}, the resulting measurement (shot) noise
would be $\approx 10^{-4} rad/\sqrt{Hz}$.  Of course this is not as
small as the shot noise at the photo detector of an optical
interferometer (in the case of LISA, for instance, this is $\approx
10^{-6} rad/\sqrt{Hz}$) simply because the typical number of photons
impinging on a photo detector is easily larger than the number of
atoms in an AI.

In order to compensate for the low number of atoms, AIs with large
momentum transfer (LMT) have been proposed
\cite{Dimopoulos_etal2008,Mueller}.  In LMT schemes, an equivalent of
multiple photon transitions between the same pair of atomic energy
levels is performed, resulting in perhaps as much as 1000-photon
momentum transfer \cite{Dimopoulos_etal2008}. This increases the
effective laser wave number to $k_{eff} \equiv Nk$ (or $\lambda_{eff}
= \lambda / N$), quite analogous to a many-bounce laser
interferometer.  The measurement sensitivity is increased without a
higher phase resolution. However, in order to accumulate the photon
momentum transfer to atoms, counter-propagating laser beams are used,
which would unfortunately break our one-way laser noise cancellation
scheme.  In order to improve the AI sensitivity by implementing some
alternative LMT scheme (and still preserve the ability of canceling
the laser phase noise) one would have to consider high-order
multi-photon transitions in an atomic or even nuclear system
characterized by large energy separations. The feasibility of such
implementation remains to be investigated.

\section{Conclusion}
\label{SecV}

The idea of using atoms as proof masses for a gravitational wave
detector is certainly very interesting, and deserves further studies.
We have presented the design concept of a one-way laser interferometer
detector of gravitational radiation that treats atoms both as
proof-masses as well as ultra-stable clocks. Our discussion has in
fact focused on the perspective of using the atomic internal
oscillations as ideal local reference clocks. The salient feature of
the one-way phase measurement configuration we have discussed is the
exact cancellation of the laser phase noise by applying TDI to the AIs
data.  Although the detection sensitivity of our detector design might
be limited by the available signal to noise ratio due to the atom shot
noise, moderate sensitivity gains might be obtainable by using
multi-photon transition. We plan to further study in a forthcoming
article this problem together with a detailed noise analysis of the
detector design presented here.

\begin{acknowledgements}
  The authors thank Professor H. Mueller for valuable discussions on
  atom interferometers and especially on the subject of LMT and
  multi-photon transitions. This research was performed at the Jet
  Propulsion Laboratory, California Institute of Technology, under
  contract with the National Aeronautics and Space Administration.(c)
  2008 California Institute of Technology. Government sponsorship
  acknowledged.
\end{acknowledgements}

\end{document}